\documentclass{midl} 

\usepackage[utf8]{inputenc} 
\usepackage[T1]{fontenc}    
\usepackage{hyperref}       
\usepackage{booktabs}       
\usepackage{amsfonts}       
\usepackage{nicefrac}       
\usepackage{microtype}      
\usepackage{lipsum}		
\usepackage{graphicx}
\usepackage{framed,multirow}
\usepackage{array}
\usepackage{enumitem}

\usepackage{latexsym}
\usepackage{caption} 

\usepackage{xcolor}

\definecolor{newcolor}{rgb}{.8,.349,.1}

\title[DL-based MO DIR]{Multi-Objective Learning for Deformable Image Registration}

\midlauthor{\Name{Monika Grewal\nametag{$^{1}$}} \Email{monika.grewal@cwi.nl}\\
\Name{Henrike Westerveld\nametag{$^{2}$}} \Email{g.westerveld@erasmusmc.nl}\\
\Name{Peter A. N. Bosman\nametag{$^{1, 3}$}} \Email{peter.bosman@cwi.nl}\\
\Name{Tanja Alderliesten\nametag{$^{4}$}} \Email{t.alderliesten@lumc.nl}\\
\addr $^{1}$ Centrum Wiskunde \& Informatica, 1098 XG, Amsterdam, The Netherlands \\
\addr $^{2}$ Erasmus University Medical Center, 3015 GD, Rotterdam, The Netherlands \\
\addr $^{3}$ Delft University of Technology, 2628 CD, Delft, The Netherlands\\
\addr $^{4}$ Leiden University Medical Center, 2333 ZC, Leiden, The Netherlands
}

\begin{document}

\maketitle

\begin{abstract}
Deformable image registration (DIR) involves optimization of multiple conflicting objectives, however, not many existing DIR algorithms are multi-objective (MO). Further, while there has been progress in the design of deep learning algorithms for DIR, there is no work in the direction of MO DIR using deep learning. In this paper, we fill this gap by combining a recently proposed approach for MO training of neural networks with a well-known deep neural network for DIR and create a deep learning based MO DIR approach. We evaluate the proposed approach for DIR of pelvic magnetic resonance imaging (MRI) scans. We experimentally demonstrate that the proposed MO DIR approach -- providing multiple registration outputs for each patient that each correspond to a different trade-off between the objectives -- has additional desirable properties from a clinical use point-of-view as compared to providing a single DIR output. The experiments also show that the proposed MO DIR approach provides a better spread of DIR outputs across the entire trade-off front than simply training multiple neural networks with weights for each objective sampled from a grid of possible values.
\end{abstract}


\section{Introduction}
\label{sec:introduction}
Deformable image registration (DIR) refers to the task of finding a non-linear transformation that aligns two images. The non-linear transformation is characterized by a deformation vector field (DVF), that maps each location in the target image (also referred to as fixed or reference image) to a location in the source image (also referred to as moving image). The source image is then warped by resampling from the mapped locations. Some of the potential applications of DIR in medical imaging are dose accumulation in radiation treatment, contour propagation, tumor growth tracking, and creating a digital atlas \cite{Mohammadi2019, Rigaud2019, ZHAO2022, salehi2022deep}.

DIR involves optimization of a parameterized DVF to maximize the similarity between two images. However, optimizing only for maximizing image similarity may yield a highly irregular or sometimes physically implausible DVF due to model overfitting. Therefore, an additional objective penalizing irregularity in the DVF is often used, which inherently conflicts with the objective of maximizing image similarity \cite{Li2018, voxelmorph, de2019deep}. Further, an additional guidance objective (either maximizing the similarity between organ contours or minimizing the distance between corresponding landmarks) is often utilized in challenging DIR problems \cite{voxelmorph, hering2021cnn}. Intuitively, improvement in the additional guidance objective should always lead to improvement in the image similarity objective. However, in practice, the additional guidance objective may still conflict with the image similarity objective. This is often caused when the optimization gets overfitted to the regions where additional guidance is provided, deteriorating performance in other image regions \cite{voxelmorph}. Another cause for conflict between the image similarity objective with the additional guidance can be the uncertainty in the additional guidance, which, in turn, could be caused by either inter/intra-observer variance in case of manual annotation or modeling error in case of automatic generation of additional guidance. Therefore, DIR is essentially a multi-objective (MO) problem \cite{deb2016multi}, which involves two or more conflicting objectives.

Therefore, fundamentally an MO approach is appropriate for DIR, where multiple DIR outputs corresponding to a diverse range of trade-offs between the conflicting objectives are provided to the clinicians to a posteriori choose the best solution. Although the notion of DIR being multi-objective is well accepted and discussed, not many DIR approaches have been developed with this perspective. \citet{alderliesten2015getting} provided a proof-of-concept study for MO DIR of 2D images. \citet{pirpinia2017feasibility} used an evolutionary algorithm to tune the corresponding weights of different objectives for each 3D breast MRI pair and run single objective DIR multiple times. \citet{nakane2022image} formulated DIR as MO problem by partitioning the template image into several overlapping regions. \citet{AndreadisMOREA} presented the first integral approach to MO DIR that could be used for 3D volumetric scans using an MO optimization algorithm.

With the advent of deep learning in the past few years, multiple deep learning based DIR approaches have been proposed \cite{voxelmorph, BobDeVos2017, Li2018, Li2022, salehi2022deep, Rigaud2019}, which provide the possibility to predict the DVF for an entire volumetric scan within seconds. However, to the best of our knowledge, there is no work done in the direction of MO DIR using deep learning. In this paper, we fill this gap and provide a novel approach for MO DIR using deep learning. To this end, we employed a well-known deep neural network for DIR, VoxelMorph \cite{voxelmorph}, and combined it with a recently proposed technique for training neural networks multi-objectively \cite{deist2023multi}. Our main contributions are the following:
\begin{itemize}
    \item We develop a deep learning based approach for MO DIR so that multiple DIR outputs corresponding to different trade-offs between multiple objectives can be presented to the clinical experts for a posteriori decision-making.
    \item We develop a parameter-efficient version of the previously introduced deep learning approach for MO DIR. With our proposed strategy, extending any encoder-decoder style DIR network to its MO version is straightforward.
    \item We demonstrate MO DIR for a challenging real-world registration task: DIR of female pelvic magnetic resonance imaging (MRI) scans and highlight its potential benefits.
\end{itemize}

\section{Approach}
\label{sec:approach}

We first provide a brief background on the concepts of MO optimization that we apply to deep learning based DIR. MO optimization refers to minimizing\footnote{In this paper, we assume minimization as objectives correspond to losses in deep learning.} a vector of $n$ objectives simultaneously. The goal is to find a set (often referred to as `approximation set') of $p$ solutions that are both close to as well as diversely-spread along the Pareto front -- the set of all Pareto optimal solutions in objective space. A solution is Pareto optimal if none of the objectives can be improved without a simultaneous detriment in performance in at least one of the other objectives \cite{van2000multiobjective}.

\begin{figure}
    \centering
    \includegraphics[width=0.8\textwidth]{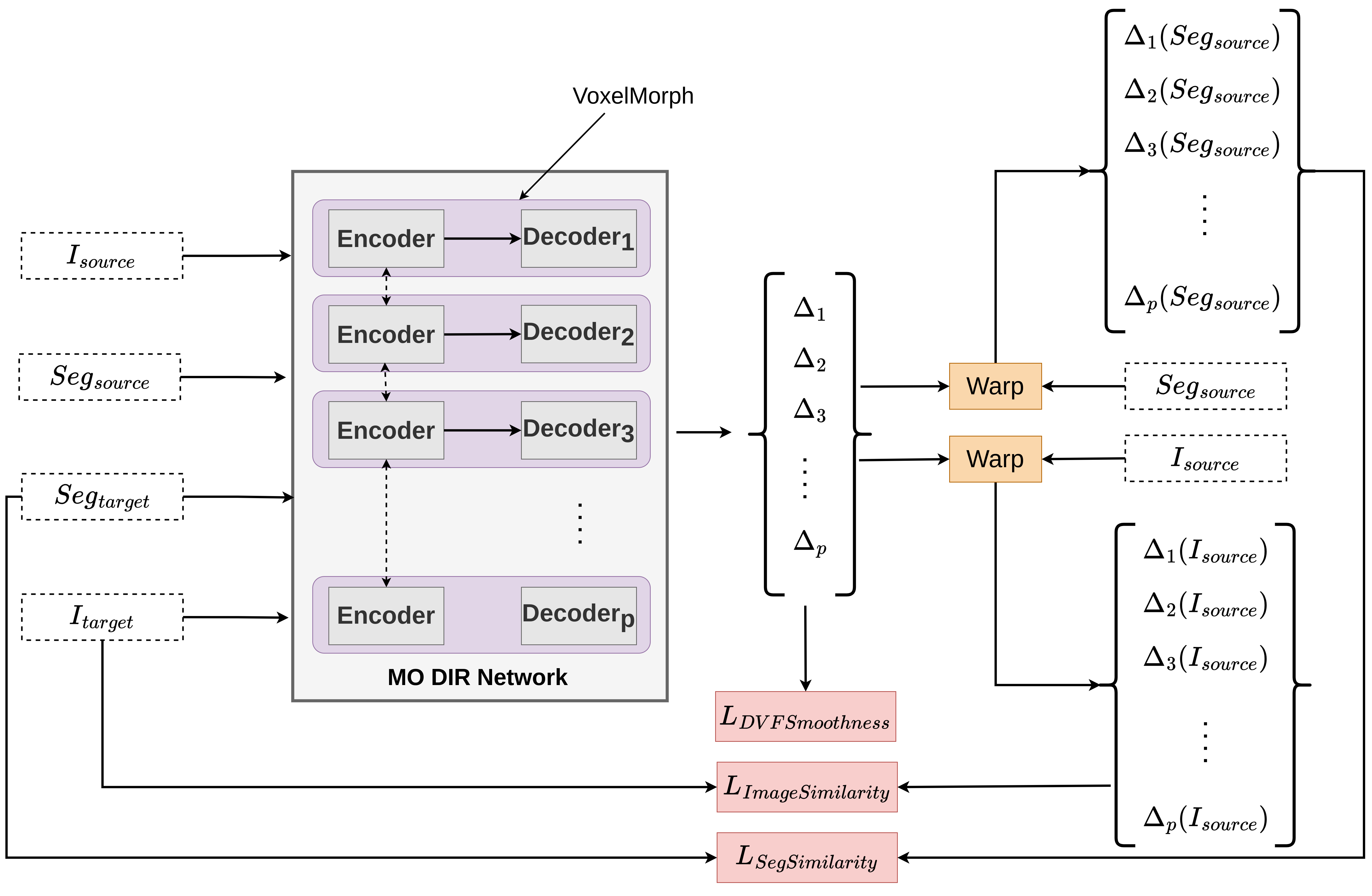}
    \caption{Illustration of the proposed deep learning based MO DIR approach. $I_{source}$: source image, $I_{target}$: target image, $Seg_{source}$ and $Seg_{target}$: organ segmentation masks for source and target image, respectively. The weights of the encoder are shared among $p$ DIR networks, which output $p$ DVFs ($\Delta_1$, $\Delta_2$, ..., $\Delta_p$) to warp $I_{source}$ and $Seg_{source}$. The network is trained to simultaneously minimize $p$ loss vectors $[L_{ImageSimilarity}, L_{DVFSmoothness}, L_{SegSimilarity}]$ using MO learning.}
    \label{fig:approach}
\end{figure}

Our deep learning based MO DIR implementation consists of a DIR network within the MO learning framework proposed in \citet{deist2023multi}. We selected VoxelMorph \cite{voxelmorph} for DIR because it is a well-known neural network for DIR. VoxelMorph uses an encoder-decoder style neural network for predicting a DVF, which is a basis for many deep learning based DIR approaches proposed afterwards. We selected the MO learning framework proposed in \citet{deist2023multi} for two reasons: a) it achieves MO training of neural networks through hypervolume (HV) maximization - a process that inherently ensures Pareto optimality\footnote{If HV is maximal, all the solutions are Pareto optimal.} and diversity between the solutions, b) it is the only MO approach that allows training neural networks multi-objectively without a priori knowledge of the exact preference between different objectives. It should be noted that the latter is crucial in the task of DIR. This is because earlier literature suggests that the exact preference between different objectives may be different between different image pairs, which may only be known a posteriori after inspecting multiple solutions \cite{pirpinia2017feasibility}.

In this paper, we aim to minimize $p$ loss vectors (corresponding to $p$ solutions or DIR outputs in the approximation set), each comprising of three losses: $L_{ImageSimilarity}$, $L_{DVFSmoothness}$, and $L_{SegSimilarity}$. Here, for $L_{ImageSimilarity}$, we used normalized cross-correlation loss. $L_{DVFSmoothness}$ is the squared sum of spatial gradients of the predicted DVF in all directions, and $L_{SegSimilarity}$ is the Dice loss between the fixed image's organ mask and the moving image's organ mask warped by the predicted DVF (refer to \cite{voxelmorph} for details). In the original formulation of MO learning in \citet{deist2023multi}, $p$ neural networks are required corresponding to $p$ solutions in the approximation set. Due to the memory intensive nature of training a 3D DIR network, this poses a challenge due to limited GPU memory. To tackle this, we modified the original implementation by sharing the weights of the encoder between $p$ DIR networks\footnote{In our preliminary experiments (shown in Appendix \ref{appendix: parameter sharing} Figure \ref{fig:effect of parameter sharing}), we observed that parameter sharing in the encoder causes a negligible decrease in the diversity of the predictions while increasing the parameter efficiency by 17\% for $p = 5$.} as shown in Figure \ref{fig:approach}. The DIR network predicts $p$ DIR outputs (DVFs). This is followed by calculation of $p$ loss vectors, which are used in the MO learning framework. The parameters of the DIR network are updated using a dynamic loss formulation, that, for each DIR output is defined as: 
\begin{multline}
\label{eq:dynamic_loss}
    L^i = w^i_1 L_{ImageSimilarity} + w^i_2 L_{DVFSmoothness} + w^i_3 L_{SegSimilarity} \quad \forall i\in\{1,\dots,p\}
\end{multline}
Where, the weights $w^i_1, w^i_2, w^i_3$ are calculated in each iteration using HV maximization. This ensures that at the end of the training the DIR outputs (that are used to calculate the $p$ loss vectors) are close to, and diversely distributed along the Pareto front of the three objectives.



\subsection{Data}
We retrospectively used data from cervical cancer patients who received brachytherapy treatment at Leiden University Medical Center (LUMC), The Netherlands. We received 136 MRI scan pairs (along with associated contours generated for clinical use of four organs at risk: bladder, bowel bag, rectum, and sigmoid) corresponding to two fractions of brachytherapy treatment in anonymized form after approval from the medical ethics committee. The original resolution of the MRI scans was 0.5 mm $\times$ 0.5 mm $\times$ 4 mm. We resampled the MRI scans to a resolution of 1 mm $\times$ 1 mm $\times$ 1 mm and used randomly cropped patches of size 192 $\times$ 192 $\times$ 32 as an input to the neural network. We separated the scans at patient level based on their chronological order of acquisition into train and validation (126 scan pairs), and test (10 scan pairs) splits. On the test scans, a radiation therapy technologist annotated 23 anatomical landmarks (details in Appendix \ref{appendix: landmarks list}), which were selected by a radiation oncologist on the basis of their importance in brachytherapy treatment for cervical cancer patients. The placement of landmarks was cross-checked by another radiation oncologist.

\section{Experiments and Results}
We implemented\footnote{The implementation will be made publicly available upon publication.} our proposed approach using Python and PyTorch. The training hyperparameters were: number of solutions $p$ = 27, initialization = Kaiming He, optimizer = Adam, learning rate (lr) = $1e^{-4}$, number of training iterations = 20K, reference point for HV calculation = (1, 1, 1) (details in Appendix \ref{appendix: reference point}). To assess the DIR performance, we calculated target registration errors (TREs) of the 23 manually annotated landmarks by transforming the landmarks in the target image with the predicted DVF and calculating the Euclidean distance with the corresponding landmarks in the source image. We also calculated the percentage of voxels with a negative determinant of the spatial Jacobian of the DVF, as an indication of folding in the transformation.

\begin{figure}
    \centering
    \includegraphics[width=\textwidth]{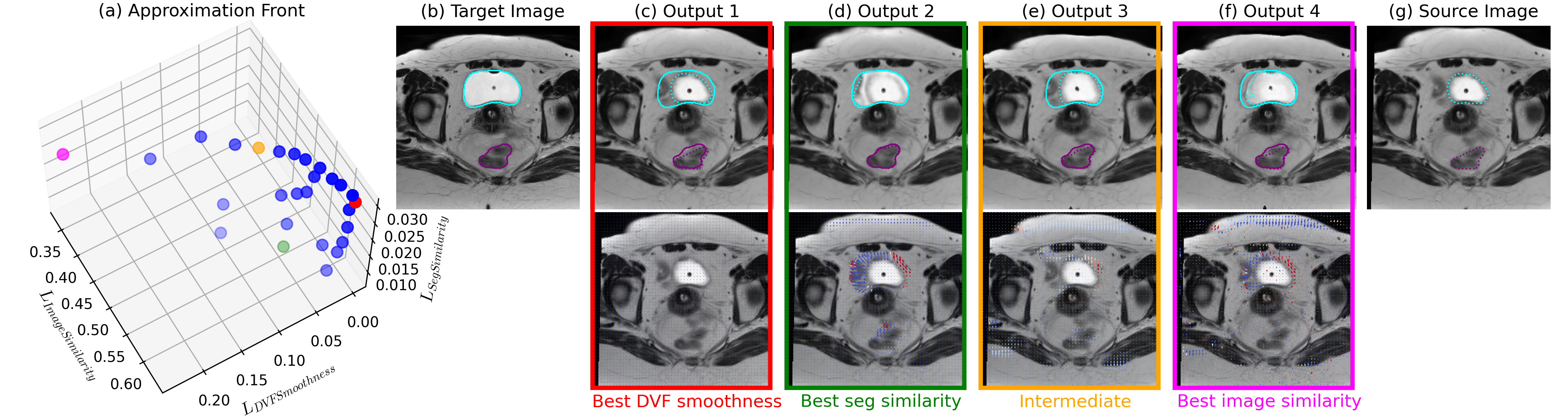}
    \caption{(a) Approximation set consisting of 27 solutions, each corresponding to a different trade-off between the 3 loss functions. (b) and (g): A transverse slice from the target and source image, respectively. (c) - (f): Warped images (top row) and DVFs overlaid on the source image (bottom row) corresponding to four solutions (highlighted in color matching with the image frame) in the set. The direction and scale of the arrows represent the displacement vector in the x-y plane, and the color (contrasting for cranial vs. caudal motion) of the arrows represent the displacement along the z-direction. Bladder and rectum contours are shown in cyan and magenta colors, respectively on the images.}
    \label{fig:demonstrate}
\end{figure}
\begin{figure}[h]
    \centering
    \includegraphics[width=0.8\textwidth]{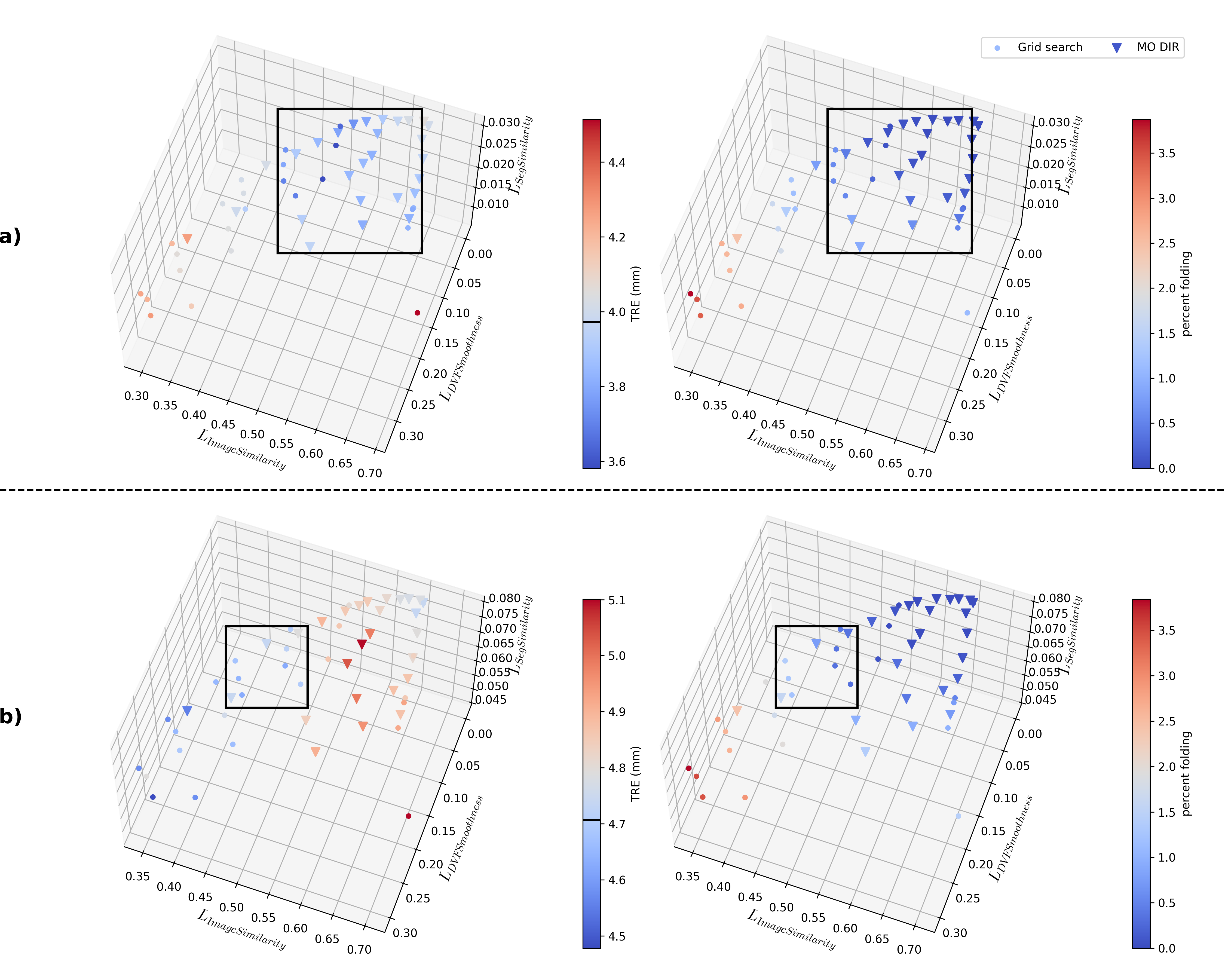}
    \caption{Approximation sets obtained for two representative test scan pairs in (a) and (b). The colors of the points represent the TRE values in mm (left), and percent folding (right). Lower values represented by blue tones are better. The TRE before DIR is represented by a black line on the TRE colorbar. Black boxes indicate the likely desired regions.}
    \label{fig:LS vs. HV}
\end{figure}

\subsection{Comparison of MO DIR with Single DIR Output}
Contrary to traditional DIR, in MO DIR, the decision maker is provided with multiple DIR solutions. This is demonstrated in Figure \ref{fig:demonstrate} (a). The figure shows that there are multiple possible ways to align the two images, which represent different trade-offs between the objectives of interest. For example in solution 2 (green), the bladder contour of the transformed source image seems perfectly aligned with the bladder contour of the target image, but there seems to be misalignment in image intensities in the top part of the image. Similarly, the transformed source image from solution 4 seems more similar to the target image as compared to the transformed source image from solution 3, however, the DVF from solution 3 seems more smooth than the DVF from solution 4. In this scenario, the decision maker (clinician) can evaluate different trade-offs and select a DIR solution specific to this image pair. Such an MO scenario also allows incorporating additional clinical criteria into decision making e.g., alignment of the bladder may be more important than the alignment of the rectum in a specific clinical use case.

In Figure \ref{fig:LS vs. HV} (a), we show the obtained approximation set for one of the test scan pairs. It is apparent that the desired solutions based on lower TRE values (solutions in blue color in the left image) are different from the desired solutions based on lower percent folding (solutions in blue color in the right image) in the transformation. There is a small subset of solutions (black boxes in Figure \ref{fig:LS vs. HV}) for which both the TRE and percent folding are in the lower range of obtained values. Further, the range of trade-offs represented by these subsets of solutions is different for the two scan pairs illustrated in Figure \ref{fig:LS vs. HV} (a) and Figure \ref{fig:LS vs. HV} (b). This demonstrates how a single DIR solution corresponding to a certain trade-off between objectives may perform better on one performance metric, but not on another performance metric. Similarly, it may provide best balance between performance metrics for one scan pair, but not for another. On the other hand, in MO DIR, a clinician can evaluate multiple possibilities corresponding to different trade-offs, separately for each scan pair, and make an informed decision.


\begin{table}[]
\centering
\caption{Minimum mean TRE of 23 anatomical landmarks in mm, associated folding in \% for grid search and MO DIR, respectively for each test scan pair. Mean $\pm$ standard deviation from 5 models from 5-fold cross-validation is reported.}
\begin{tabular}{@{}lc|cc|cc@{}}
\toprule
\multirow{2}{*}{Test scan}    & \multirow{2}{*}{TRE before} & \multicolumn{2}{c}{\textbf{Grid Search}} & \multicolumn{2}{c}{\textbf{MO DIR}} \\ \cmidrule(l){3-6} 
     && TRE                  & \% folding                       & TRE                   & \% folding                       \\ \midrule
1 & 3.97 & 3.63 $\pm$ 0.04 & 0.29 $\pm$ 0.19  & 3.74 $\pm$ 0.03, & 0.05 $\pm$ 0.03\\
2 & 4.71 & 4.53 $\pm$ 0.11 & 3.45 $\pm$ 0.38  & 4.66 $\pm$ 0.07, & 2.00 $\pm$ 1.23\\ 
3 & 8.21 & 8.04 $\pm$ 0.10 & 1.33 $\pm$ 1.18  & 8.12 $\pm$ 0.06, & 1.07 $\pm$ 1.64\\ 
4 & 9.07 & 8.18 $\pm$ 0.07 & 0.12 $\pm$ 0.15  & 8.58 $\pm$ 0.17, & 0.47 $\pm$ 0.39\\ 
5 & 4.46 & 4.01 $\pm$ 0.06 & 0.80 $\pm$ 0.96  & 4.08 $\pm$ 0.07, & 1.36 $\pm$ 1.01\\ 
6 & 5.55 & 4.52 $\pm$ 0.09 & 1.31 $\pm$ 0.17  & 4.69 $\pm$ 0.09, & 0.76 $\pm$ 0.32\\ 
7 & 5.99 & 5.90 $\pm$ 0.03 & 0.26 $\pm$ 0.18  & 5.93 $\pm$ 0.02, & 0.29 $\pm$ 0.13\\ 
8 & 4.39 & 3.96 $\pm$ 0.05 & 2.72 $\pm$ 0.88  & 4.06 $\pm$ 0.05, & 1.72 $\pm$ 1.31\\ 
9 & 5.73 & 5.06 $\pm$ 0.06 & 0.87 $\pm$ 0.24  & 5.24 $\pm$ 0.13, & 0.82 $\pm$ 0.97\\ 
10 & 3.80 & 3.72 $\pm$ 0.03 & 0.20 $\pm$ 0.28  & 3.70 $\pm$ 0.03, & 0.11 $\pm$ 0.13\\ \bottomrule
\end{tabular}
\label{tab:descriptives}
\end{table}

\subsection{Comparison of Proposed MO DIR with Grid Search}
Traditionally, a set of neural networks are trained with a weighted loss formulation with weights for each loss sampled from a grid of possible values. Finally, one of these neural networks is then selected based on the performance on validation data. To simulate the MO DIR set up with grid search, we trained the different heads of our MO DIR neural network with weights corresponding to different points in grid search. We used 27 grid points by enumerating over all the possible combinations for $w_1 \in \{0, 0.5, 1\}$, $w_2 \in \{0, 0.1, 0.5, 1\}$, and $w_3 \in \{0, 0.5, 1\}$ and omitting redundant (e.g., $\{0, 0.5, 0.5\}$ and $\{0, 0.1, 0.1\}$) and less meaningful (e.g., $\{0, 1, 0\}$) combinations. It should be noted that this process of grid search is already slightly better than naive grid search. Moreover, it incorporates the domain knowledge that training only with deformation smoothness loss is not meaningful - something which is not required for HV maximization based MO DIR approach.

The obtained approximation set from these solutions is shown with filled circles in Figure \ref{fig:LS vs. HV}. Figure \ref{fig:LS vs. HV} shows that the obtained solutions from grid search are clustered in certain regions and do not provide diversity in trade-offs between objectives. This demonstrates the difficulty in tuning the weights of different objectives to achieve a diverse range of trade-offs with grid search. On the other hand, with the proposed MO DIR approach, the tuning for both diversity as well as proximity to the Pareto front is achieved in a straightforward manner. Further, in Table \ref{tab:descriptives}, the best TRE values along with the corresponding percent folding in the transformed images obtained by grid search solutions and MO DIR solutions is shown for each scan pair. The table indicates that both grid search and the proposed MO DIR find quantitatively similar trade-offs between the best TRE values and associated image folding. It is important to remark here that TRE is a sparse metric for assessing the DIR performance. Moreover, the image folding is an aggregate metric over entire scan. As such, it is difficult to compare between different trade-offs or derive clinical conclusions without inspecting the underlying DVFs.


\section{Conclusions and Discussion}
We propose the first deep learning approach for MO DIR, which provides multiple DIR solutions diversely spread across the trade-off front between conflicting objectives. With such an approach, clinicians can evaluate the provided DIR solutions according to patient-specific clinical criteria and select the most suitable trade-off. We experimentally demonstrated the added value of MO DIR as compared to providing a single DIR output in terms of providing insights, and capability to select the most suitable DIR solution for each patient case, which potentially increases the chances of clinical adoption. We also demonstrated that the MO DIR setting is more efficient than a grid search for weights for each objective in terms of spread of solutions across the approximation front. 

Although the potential utility of deep learning based MO DIR is evident from experimental results, the presented work is still only a proof-of-concept. It has certain limitations and because the concept of MO learning is new to the field of DIR, there exist some open questions. Some of the limitations, open questions, and possible future research directions are as follows:

\begin{itemize}
    \item HV maximization provides a straightforward way to distribute the solutions diversely on the approximation front without requiring any manual tuning. A downside of this is that it is difficult to find differently distributed points, for example, if the solutions are desired to be biased towards one objective. In future work, it would be interesting to investigate the use of the weighted HV \cite{weighted_hypervolume} metric in MO DIR to steer the solutions to a desired region. It is also important to investigate which part of the approximation front is more desired by involving clinicians as a posteriori decision-makers.
    \item In Figures \ref{fig:demonstrate} and \ref{fig:LS vs. HV}, the solutions seem more clustered in the corner where $L_{ImageSimilarity}$ and $L_{SegSimilarity}$ are large and $L_{DVFSmoothness}$ is small. This could be because this corner of the front is easy to achieve due to no or little deformation, or because of the corresponding shape of and local density along the Pareto front. It is known that setting the reference point differently can impact this \cite{ishibuchi2018specify} (also see Appendix \ref{appendix: reference point}). It will be interesting to investigate this further in future.
    \item In our proof-of-principle, we made certain choices e.g., number of objectives, number of solutions in the approximation set, type of additional guidance, type of neural network for DIR, in an effort to create a baseline deep learning based MO DIR approach. That said, the current approach leaves multiple improvement possibilities open in order to realize the complete potential of the MO perspective for DIR. For example, it can be improved by using a more sophisticated neural network for DIR, multi-resolution registration, constraints on tissue types, and diffeomorphism.
    \item The presented MO DIR work provides more insights than traditional approaches by showcasing the trade-offs between different objectives and how these trade-offs differ between scan pairs. However, the objectives are still average values per pair of scans. Practically, the DIR performance will likely not be uniform across the entire scan. Additionally, it is possible that clinically a solution in the vicinity of a provided discrete solution on the approximation front is more desired. It is therefore essential to research in the direction of intuitively visualizing the DVFs and navigating across (and in the local neighborhood of) different solutions.
\end{itemize}

\midlacknowledgments{We thank W. Visser-Groot and S.M. de Boer (Dept. of Radiation Oncology, LUMC, Leiden, NL) for their contributions to
this study. This research is part of the research programme Open
Technology Programme with project number 15586, which is financed by the Dutch Research Council (NWO), Elekta, and Xomnia.
Further, the work is co-funded by the public-private partnership
allowance for top consortia for knowledge and innovation (TKIs)
from the Dutch Ministry of Economic Affairs.}

\bibliography{dir-refs}

\begin{thebibliography}{20}
\providecommand{\natexlab}[1]{#1}
\providecommand{\url}[1]{\texttt{#1}}
\expandafter\ifx\csname urlstyle\endcsname\relax
  \providecommand{\doi}[1]{doi: #1}\else
  \providecommand{\doi}{doi: \begingroup \urlstyle{rm}\Url}\fi

\bibitem[Alderliesten et~al.(2015)Alderliesten, Bosman, and Bel]{alderliesten2015getting}
Tanja Alderliesten, Peter A.~N. Bosman, and Arjan Bel.
\newblock Getting the most out of additional guidance information in deformable image registration by leveraging multi-objective optimization.
\newblock In \emph{Medical Imaging 2015: Image Processing}, volume 9413 of \emph{Proc. SPIE}, page 94131R. International Society for Optics and Photonics, 2015.

\bibitem[Andreadis et~al.(2023)Andreadis, Bosman, and Alderliesten]{AndreadisMOREA}
Georgios Andreadis, Peter~A.N. Bosman, and Tanja Alderliesten.
\newblock {MOREA}: A {GPU}-accelerated evolutionary algorithm for multi-objective deformable registration of 3d medical images.
\newblock In \emph{Proceedings of the Genetic and Evolutionary Computation Conference}, GECCO '23, page 1294–1302, New York, NY, USA, 2023. Association for Computing Machinery.
\newblock ISBN 9798400701191.
\newblock \doi{10.1145/3583131.3590414}.
\newblock URL \url{https://doi.org/10.1145/3583131.3590414}.

\bibitem[Balakrishnan et~al.(2019)Balakrishnan, Zhao, Sabuncu, Guttag, and Dalca]{voxelmorph}
Guha Balakrishnan, Amy Zhao, Mert~R. Sabuncu, John Guttag, and Adrian~V. Dalca.
\newblock {VoxelMorph}: A learning framework for deformable medical image registration.
\newblock \emph{IEEE Transactions on Medical Imaging}, 38\penalty0 (8):\penalty0 1788--1800, 2019.
\newblock \doi{10.1109/TMI.2019.2897538}.

\bibitem[Bosman(2011)]{bosman2011gradients}
Peter~AN Bosman.
\newblock On gradients and hybrid evolutionary algorithms for real-valued multiobjective optimization.
\newblock \emph{IEEE Transactions on Evolutionary Computation}, 16\penalty0 (1):\penalty0 51--69, 2011.

\bibitem[de~Vos et~al.(2017)de~Vos, Berendsen, Viergever, Staring, and I{\v{s}}gum]{BobDeVos2017}
Bob~D. de~Vos, Floris~F. Berendsen, Max~A. Viergever, Marius Staring, and Ivana I{\v{s}}gum.
\newblock End-to-end unsupervised deformable image registration with a convolutional neural network.
\newblock In M.~Jorge Cardoso, Tal Arbel, Gustavo Carneiro, Tanveer Syeda-Mahmood, Jo{\~a}o Manuel~R.S. Tavares, Mehdi Moradi, Andrew Bradley, Hayit Greenspan, Jo{\~a}o~Paulo Papa, Anant Madabhushi, Jacinto~C. Nascimento, Jaime~S. Cardoso, Vasileios Belagiannis, and Zhi Lu, editors, \emph{Deep Learning in Medical Image Analysis and Multimodal Learning for Clinical Decision Support}, pages 204--212, Cham, 2017. Springer International Publishing.
\newblock ISBN 978-3-319-67558-9.

\bibitem[De~Vos et~al.(2019)De~Vos, Berendsen, Viergever, Sokooti, Staring, and I{\v{s}}gum]{de2019deep}
Bob~D De~Vos, Floris~F Berendsen, Max~A Viergever, Hessam Sokooti, Marius Staring, and Ivana I{\v{s}}gum.
\newblock A deep learning framework for unsupervised affine and deformable image registration.
\newblock \emph{Medical image analysis}, 52:\penalty0 128--143, 2019.

\bibitem[Deb et~al.(2016)Deb, Sindhya, and Hakanen]{deb2016multi}
Kalyanmoy Deb, Karthik Sindhya, and Jussi Hakanen.
\newblock Multi-objective optimization.
\newblock In \emph{Decision sciences}, pages 161--200. CRC Press, 2016.

\bibitem[Deist et~al.(2023)Deist, Grewal, Dankers, Alderliesten, and Bosman]{deist2023multi}
Timo~M Deist, Monika Grewal, Frank~JWM Dankers, Tanja Alderliesten, and Peter~AN Bosman.
\newblock Multi-objective learning using {HV} maximization.
\newblock In \emph{International Conference on Evolutionary Multi-Criterion Optimization}, pages 103--117. Springer, 2023.

\bibitem[Hering et~al.(2021)Hering, H{\"a}ger, Moltz, Lessmann, Heldmann, and van Ginneken]{hering2021cnn}
Alessa Hering, Stephanie H{\"a}ger, Jan Moltz, Nikolas Lessmann, Stefan Heldmann, and Bram van Ginneken.
\newblock Cnn-based lung ct registration with multiple anatomical constraints.
\newblock \emph{Medical Image Analysis}, 72:\penalty0 102139, 2021.

\bibitem[Ishibuchi et~al.(2018)Ishibuchi, Imada, Setoguchi, and Nojima]{ishibuchi2018specify}
Hisao Ishibuchi, Ryo Imada, Yu~Setoguchi, and Yusuke Nojima.
\newblock How to specify a reference point in hypervolume calculation for fair performance comparison.
\newblock \emph{Evolutionary Computation}, 26\penalty0 (3):\penalty0 411--440, 2018.

\bibitem[Li and Fan(2018)]{Li2018}
Hongming Li and Yong Fan.
\newblock Non-rigid image registration using self-supervised fully convolutional networks without training data.
\newblock In \emph{2018 IEEE 15th International Symposium on Biomedical Imaging (ISBI 2018)}, pages 1075--1078, 2018.
\newblock \doi{10.1109/ISBI.2018.8363757}.

\bibitem[Li et~al.(2022)Li, Fan, and for~the Alzheimer's Disease Neuroimaging~Initiative]{Li2022}
Hongming Li, Yong Fan, and for~the Alzheimer's Disease Neuroimaging~Initiative.
\newblock {MDReg-Net}: Multi-resolution diffeomorphic image registration using fully convolutional networks with deep self-supervision.
\newblock \emph{Human Brain Mapping}, 43\penalty0 (7):\penalty0 2218--2231, 2022.
\newblock \doi{https://doi.org/10.1002/hbm.25782}.
\newblock URL \url{https://onlinelibrary.wiley.com/doi/abs/10.1002/hbm.25782}.

\bibitem[Mohammadi et~al.(2019)Mohammadi, Mahdavi, Jaberi, Siavashpour, Janani, Meigooni, and Reiazi]{Mohammadi2019}
R.~Mohammadi, S.~R. Mahdavi, R.~Jaberi, Z.~Siavashpour, L.~Janani, A.~S. Meigooni, and R.~Reiazi.
\newblock Evaluation of deformable image registration algorithm for determination of accumulated dose for brachytherapy of cervical cancer patients.
\newblock \emph{Journal of Contemporary Brachytherapy}, 11:\penalty0 469--478, 2019.

\bibitem[Nakane et~al.(2022)Nakane, Xie, and Zhang]{nakane2022image}
Takumi Nakane, Haoran Xie, and Chao Zhang.
\newblock Image deformation estimation via multiobjective optimization.
\newblock \emph{IEEE Access}, 10:\penalty0 53307--53323, 2022.

\bibitem[Pirpinia et~al.(2017)Pirpinia, Bosman, Loo, Winter-Warnars, Janssen, Scholten, Sonke, Van~Herk, and Alderliesten]{pirpinia2017feasibility}
Kleopatra Pirpinia, Peter~AN Bosman, Claudette~E Loo, Gonneke Winter-Warnars, Natasja~NY Janssen, Astrid~N Scholten, Jan-Jakob Sonke, Marcel Van~Herk, and Tanja Alderliesten.
\newblock The feasibility of manual parameter tuning for deformable breast {MR} image registration from a multi-objective optimization perspective.
\newblock \emph{Physics in Medicine \& Biology}, 62\penalty0 (14):\penalty0 5723, 2017.

\bibitem[Rigaud et~al.(2019)Rigaud, Klopp, Vedam, Venkatesan, Taku, Simon, Haigron, Crevoisier, Brock, and Cazoulat]{Rigaud2019}
B.~Rigaud, A.~Klopp, S.~Vedam, A.~Venkatesan, N.~Taku, A.~Simon, P.~Haigron, R.~De Crevoisier, K.~K. Brock, and G.~Cazoulat.
\newblock Deformable image registration for dose mapping between external beam radiotherapy and brachytherapy images of cervical cancer.
\newblock \emph{Physics in Medicine and Biology}, 64:\penalty0 115023, 2019.

\bibitem[Salehi et~al.(2022)Salehi, Vafaei~Sadr, Mahdavi, Arabi, Shiri, and Reiazi]{salehi2022deep}
Mohammad Salehi, Alireza Vafaei~Sadr, Seied~Rabi Mahdavi, Hossein Arabi, Isaac Shiri, and Reza Reiazi.
\newblock Deep learning-based non-rigid image registration for high-dose rate brachytherapy in inter-fraction cervical cancer.
\newblock \emph{Journal of Digital Imaging}, pages 1--14, 2022.

\bibitem[Van~Veldhuizen and Lamont(2000)]{van2000multiobjective}
David~A Van~Veldhuizen and Gary~B Lamont.
\newblock Multiobjective evolutionary algorithms: Analyzing the state-of-the-art.
\newblock \emph{Evolutionary computation}, 8\penalty0 (2):\penalty0 125--147, 2000.

\bibitem[Zhao et~al.(2022)Zhao, Chen, Qiu, Zhang, Liu, Zhang, Zhang, Jiang, and Wang]{ZHAO2022}
Tiandi Zhao, Yi~Chen, Bin Qiu, Jiashuang Zhang, Hao Liu, Xile Zhang, Ruilin Zhang, Ping Jiang, and Junjie Wang.
\newblock Evaluating the accumulated dose distribution of organs at risk in combined radiotherapy for cervical carcinoma based on deformable image registration.
\newblock \emph{Brachytherapy}, 2022.
\newblock ISSN 1538-4721.
\newblock \doi{https://doi.org/10.1016/j.brachy.2022.09.001}.
\newblock URL \url{https://www.sciencedirect.com/science/article/pii/S1538472122001635}.

\bibitem[Zitzler et~al.(2007)Zitzler, Brockhoff, and Thiele]{weighted_hypervolume}
Eckart Zitzler, Dimo Brockhoff, and Lothar Thiele.
\newblock The hypervolume indicator revisited: On the design of pareto-compliant indicators via weighted integration.
\newblock In Shigeru Obayashi, Kalyanmoy Deb, Carlo Poloni, Tomoyuki Hiroyasu, and Tadahiko Murata, editors, \emph{Evolutionary Multi-Criterion Optimization}, pages 862--876, Berlin, Heidelberg, 2007. Springer Berlin Heidelberg.
\newblock ISBN 978-3-540-70928-2.

\end{thebibliography}

\newpage
\appendix
\section{Effect of Parameter Sharing in the Encoder}
\label{appendix: parameter sharing}

\begin{figure}[h]
    \centering
    \includegraphics[width=\textwidth]{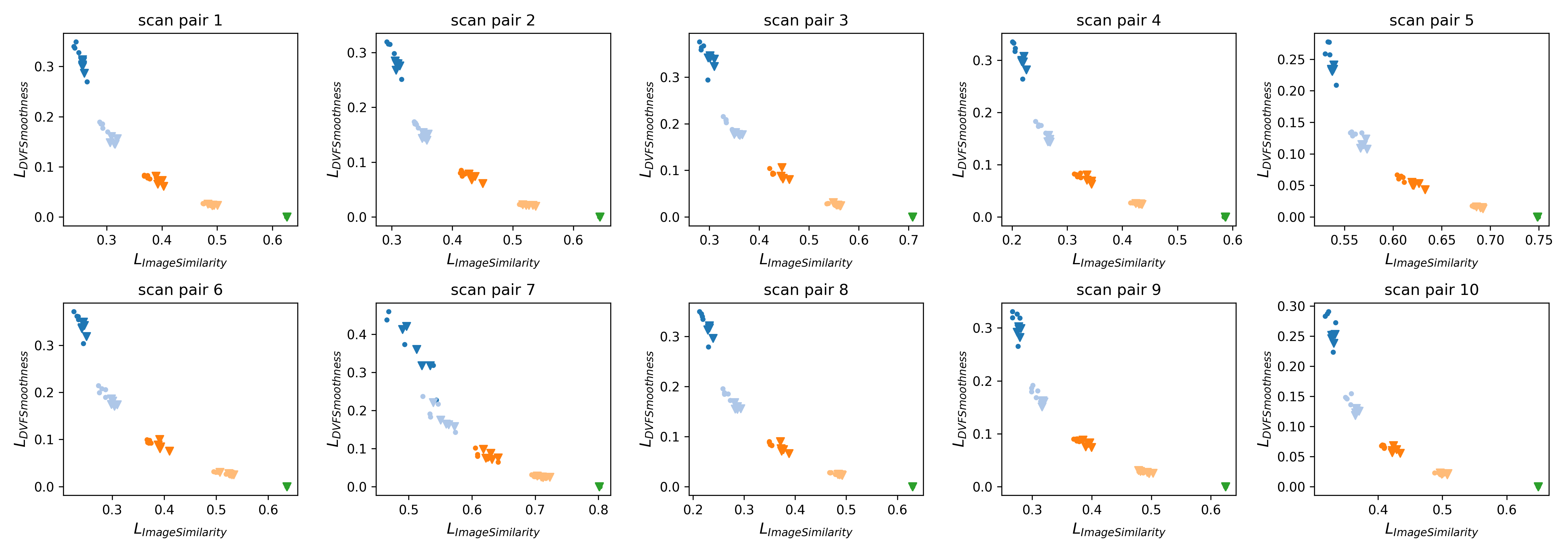}
    \caption{Effect of parameter sharing in the Encoder. filled circles: MO DIR without parameter sharing in the encoder, triangles: MO DIR with parameter sharing in the encoder. $p=5$, $n=2$. Approximation sets obtained from 5 models of 5-fold cross-validation are shown.}
    \label{fig:effect of parameter sharing}
\end{figure}

In Figure \ref{fig:effect of parameter sharing}, 5 approximation sets obtained from 5 models after 5-fold cross-validation, by training the MO DIR approach with $p=5$ for $L_{ImageSimilarity}$, and $L_{DVFSmoothness}$ losses without (filled circles) and with parameter sharing (triangles) in the encoder are shown for all the test scan pairs. The figure shows that parameter sharing does not impact the distribution of solutions on the front.

\section{Description of Landmarks}
\label{appendix: landmarks list}
\begin{itemize}[noitemsep]
    \item Internal and external urethral ostium
    \item Uterus top
    \item Cervical ostium
    \item Isthmus
    \item Intra-uterine canal top
    \item Internal anal sfincter
    \item Os coccygis
    \item Most ventral intersections of S1-S2, S2-S3, S3-S4
    \item Anterior superior border sympysis (ASBS)
    \item Posterior inferior border sympysis (PIBS)
    \item Right and Left ureteral ostium
    \item Right and left femur head
    \item Right and left acetabulum
    \item Right and left ligament rotundum
    \item Right and left entrance of uterine artery to cervix
\end{itemize}

\section{Effect of Selecting Reference Point}
\label{appendix: reference point}
\begin{figure}
    \centering
    \includegraphics[width=\textwidth]{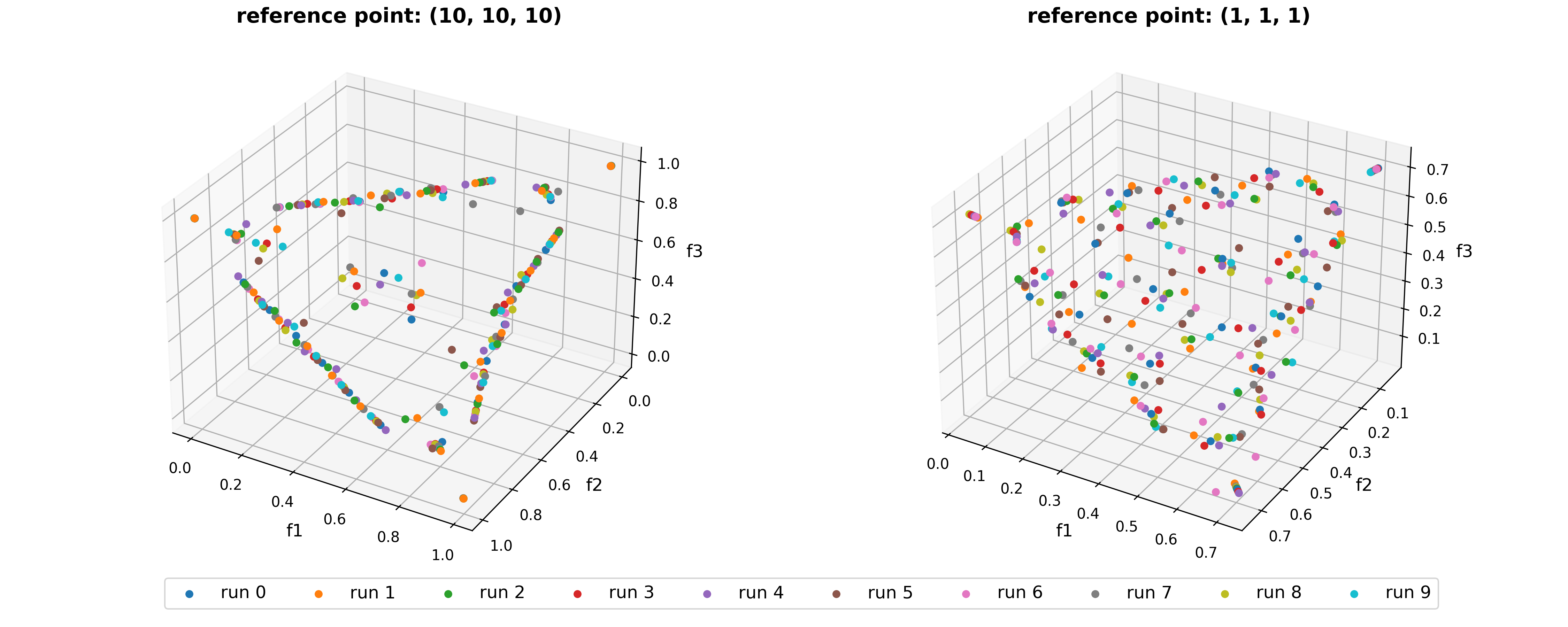}
    \caption{Effect of the location of reference point on the GenMED \cite{bosman2011gradients} benchmark problem. The Pareto front was approximated using 25 points. The solutions from 10 runs are shown for two different locations of the reference point.}
    \label{fig:effect of reference point}
\end{figure}
The calculation of the HV (and consequently its gradients) is sensitive to the choice of the reference point \cite{ishibuchi2018specify}, which, in turn, affects the spread of the solutions on the front. This is particularly the case for three or more objectives.  In Figure \ref{fig:effect of reference point}, this phenomenon is illustrated with experiments on the convex GenMED problem with three objectives \cite{bosman2011gradients}. Briefly, in the GenMED problem, the $n$ objectives (in our case, $n = 3$ i.e., f1, f2, f3 are the sum of square distances from $n$ unit vectors. When the reference point is far away, the final solutions tend to cluster on the edges of the Pareto front. The spread of the points becomes more uniform across the Pareto front when the reference point is moved closer. Based on these empirical observations, we tuned the reference point for MO DIR training. We considered the following choices: (10, 10, 10), (1, 1, 1), (1, 1, 0.2), (0.5, 1, 1) based on observing the worst loss values after training. For experiments in the paper, we selected (1, 1, 1) as the reference point because it provided well distributed points across the front based on visual inspection on validation set.

\section{Comparison between MO DIR Without and With Additional Guidance}
\begin{figure}
    \centering
    \includegraphics[width=\textwidth]{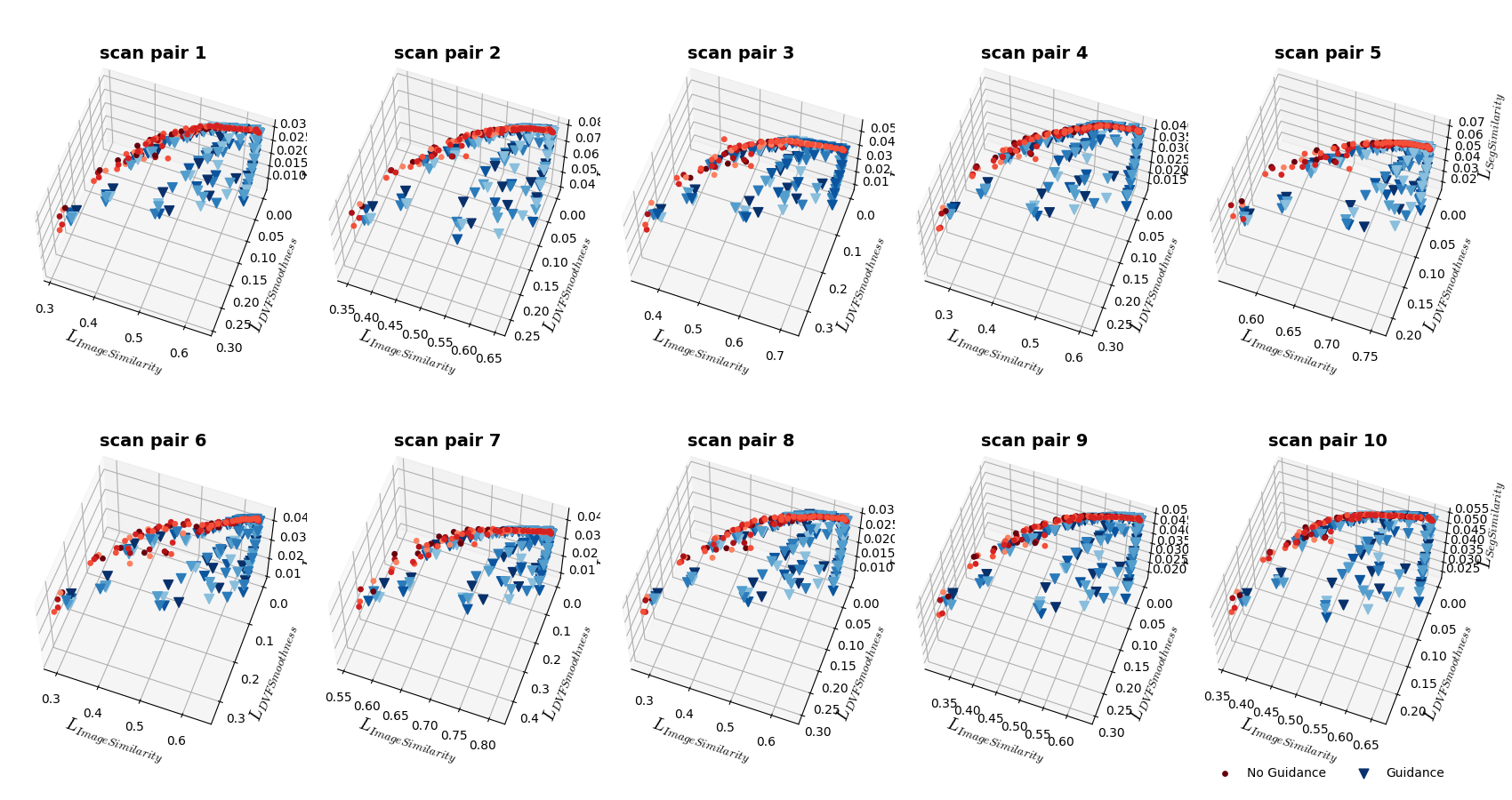}
    \caption{Effect of additional guidance. filled circles: MO DIR without additional guidance from organ contours, triangles: MO DIR with additional guidance from organ contours. $p=27$. Approximation sets obtained from 5 models of 5-fold cross-validation are shown.}
    \label{fig:effect of additional guidance}
\end{figure}

To gain insights into the effect of additional guidance from organ masks on the DIR performance, we compared the following two settings: : a) MO DIR using $L_{ImageSimilarity}$, and $L_{DVFSmoothness}$ (no additional guidance), b) MO DIR using $L_{ImageSimilarity}$, $L_{DVFSmoothness}$, and $L_{SegSimilarity}$ (additional guidance). In Figure \ref{fig:effect of additional guidance}, the obtained approximation sets on test scan pairs from both settings are shown in the objective space of $L_{ImageSimilarity}$, $L_{DVFSmoothness}$, and $L_{SegSimilarity}$. The figure shows that training MO DIR with the additional guidance from organ masks, some solutions are obtained in the region corresponding to lower $L_{SegSimilarity}$ loss but higher $L_{ImageSimilarity}$ loss. These solutions underline the conflict between $L_{ImageSimilarity}$ and $L_{SegSimilarity}$, whose nature and causes could only be known after exploring the DIR outputs corresponding to these solutions. It is worth noting that with MO DIR, such an exploratory analysis is possible and straight-forward. 

Furthermore, in Table \ref{tab:additional guidance}, the maximum mean Dice score and \% folding in the associated DVF of an approximation set is reported for each test scan pair. Similar to Figure \ref{fig:effect of additional guidance}, Table \ref{tab:additional guidance} also shows that by training DIR with additional guidance from organ masks, higher similarity between organ masks (indicated by high Dice scores) can be achieved without compromising with \% folding in the DVFs. 
It is important to state here that the best solutions in the approximation sets according to Dice score (reported in Table \ref{tab:additional guidance}) are not same as the best solutions according to TRE values (reported in Table \ref{tab:descriptives}), highlighting the nuances of evaluating a DIR outcome. Further, it is difficult to make clinically relevant performance comparisons solely based on quantitative values due to two reasons: a) mean Dice score is biased towards large organs, b) the solution corresponding to maximum Dice score may be overfitted to $L_{SegSimilarity}$ loss.

\begin{table}[]
\centering
\caption{Maximum mean percent Dice score of four organs at risk (bowel bag, bladder, rectum, and sigmoid), and associated \% folding for approximation sets obtained from MO DIR without and with guidance from organ masks, for each test scan pair. Mean $\pm$ standard deviation from 5 models from 5-fold cross-validation is reported.}
\begin{tabular}{@{}l|cc|cc@{}}
\toprule
\multirow{2}{*}{Test scan}  & \multicolumn{2}{c}{\textbf{No Guidance}} & \multicolumn{2}{c}{\textbf{Guidance}} \\ \cmidrule(l){2-5} 
     & \% Dice                  & \% folding                       & \% Dice                   & \% folding                       \\ \midrule
1 & 97.63 $\pm$ 0.04 & 1.24 $\pm$ 0.26  & 99.28 $\pm$ 0.06, & 0.77 $\pm$ 0.22\\
2 & 92.75 $\pm$ 0.09 & 1.03 $\pm$ 0.28  & 95.66 $\pm$ 0.26, & 1.38 $\pm$ 0.28\\ 
3 & 96.25 $\pm$ 0.07 & 0.64 $\pm$ 0.37  & 98.99 $\pm$ 0.10, & 0.93 $\pm$ 0.17\\ 
4 & 96.56 $\pm$ 0.04 & 0.69 $\pm$ 0.18  & 98.53 $\pm$ 0.13, & 0.87 $\pm$ 0.28\\ 
5 & 94.58 $\pm$ 0.05 & 0.03 $\pm$ 0.07  & 98.24 $\pm$ 0.09, & 0.66 $\pm$ 0.07\\ 
6 & 96.49 $\pm$ 0.13 & 1.36 $\pm$ 0.27  & 98.73 $\pm$ 0.13, & 1.00 $\pm$ 0.37\\ 
7 & 96.93 $\pm$ 0.02 & 0.93 $\pm$ 0.53  & 99.01 $\pm$ 0.11, & 0.96 $\pm$ 0.49\\ 
8 & 97.56 $\pm$ 0.09 & 0.63 $\pm$ 0.16  & 99.07 $\pm$ 0.07, & 0.64 $\pm$ 0.11\\ 
9 & 95.63 $\pm$ 0.03 & 1.89 $\pm$ 0.44  & 98.01 $\pm$ 0.11, & 1.09 $\pm$ 0.42\\ 
10 & 95.04 $\pm$ 0.01 & 0.67 $\pm$ 0.17  & 97.48 $\pm$ 0.12, & 1.02 $\pm$ 0.26\\ \bottomrule
\end{tabular}
\label{tab:additional guidance}
\end{table}

\end{document}